\documentclass[%
 reprint,
 preprintnumbers,
 amssymb,
 aps,
]{revtex4-1}

\usepackage{amsmath}
\usepackage[dvipdfmx]{graphicx}
\usepackage{dcolumn}
\usepackage{bm}
\usepackage{braket}
\usepackage{xspace}

\newcommand{\muetoee}{$\mu^-e^-\to e^-e^-$\xspace}
\usepackage{color}

\usepackage{ulem}
\usepackage{lineno}

\begin{document}

\preprint{STUPP-19-238, OCU-PHYS 502, NITEP 20}

\title{
Momentum distribution of the electron pair from the charged lepton flavor violating process $\mu^-e^-\to e^-e^-$ in muonic atoms with a polarized muon
}

\author{
Yoshitaka Kuno$^1$, Joe Sato$^2$, Toru Sato$^{3,4}$, Yuichi Uesaka$^2$, and Masato Yamanaka$^{5,6}$
}
\affiliation{
$^1$Department of Physics, Osaka University, Toyonaka, Osaka 560-0043, Japan\\
$^2$Physics Department, Saitama University, 255 Shimo-Okubo, Sakura-ku, Saitama, Saitama 338-8570, Japan\\
$^3$Research Center for Nuclear Physics (RCNP), Osaka University, Ibaraki, Osaka, 567-0047, Japan\\
$^4$J-PARC Branch, KEK Theory Center, Institute of Particle and Nuclear Studies, KEK, Tokai, Ibaraki 319-1106, Japan\\
$^5$Department of Mathematics and Physics, Osaka City University, Osaka 558-8585, Japan\\
$^6$Nambu Yoichiro Institute of Theoretical and Experimental Physics (NITEP), Osaka City University, Osaka 558-8585, Japan
}




\date{\today}

\begin{abstract}
  The \muetoee process in a muonic atom is one of the promising probes
  to study the charged lepton flavor violation (CLFV).
  We have investigated the angular distribution of electrons from the
  polarized muon of the atomic bound state.
  The parity violating asymmetric distribution of electrons
  is analyzed by using lepton wave functions under the Coulomb interaction
  of a finite nuclear charge distribution.
  It is found that the asymmetry parameters
  of electrons are very sensitive to the chiral structure of the CLFV
  interaction and the contact/photonic interaction.
  Therefore, together with the atomic number dependence of the
  decay rate studied in our previous work,
  the angular distribution of electrons from a polarized muon
  should be a very useful tool to constrain the model beyond the standard model.
\end{abstract}

\maketitle

\onecolumngrid


\section{Introduction \label{sec:Introduction}}

The charged lepton flavor violation (CLFV) is 
an excellent probe of new physics beyond the standard model (SM)
\cite{Calibbi2017} since it is highly suppressed in the SM.
The best experimental constraints of the CLFV are obtained from exotic decays of muons,
e.g., $Br\left(\mu^+\to e^+\gamma\right)<4.2\times 10^{-13}$ \cite{Baldini2016}, $Br\left(\mu^+\to e^+e^+e^-\right)<1.0\times 10^{-12}$ \cite{Bellgardt1988}, and $Br\left(\mu^-\textrm{Au}\to e^-\textrm{Au}\right)<7\times 10^{-13}$ \cite{Bertl2006}.
Next-generation experiments are planned to discover CLFV
\cite{Baldini2013,Blondel2013,COMET2018,Bartoszek2015,Nguyen2015}.

The \muetoee transition in a muonic atom was proposed as a new process to
search for the CLFV in Ref.~\cite{Koike2010}.
The \muetoee process has
interesting features complementary to the other CLFV searches.
One of its important properties is the clear signal of the process; the
total energy of two emitted electrons is equal to $m_\mu-B_\mu+m_e-B_e$,
where $B_\ell$ is the binding energy of the lepton $\ell$ in an atomic
orbit.
A discussion to search for the \muetoee process is ongoing in the COMET
Phase-I experiment at J-PARC \cite{COMET2018}.

In our recent work \cite{Uesaka2016,Uesaka2018}, careful calculations
were made for the transition rate of the \muetoee, which included the
calculation of wave functions for the bound leptons in the initial state
and the emitted electrons in the final state using a Dirac equation
with a realistic charge distribution of nuclei.
In calculating the rate of the \muetoee process, it is essential to take
into account the relativistic effects for the bound leptons and the distortion effects for the emitted electrons with the finite charge distribution.

The effective Lagrangian of the CLFV process \muetoee is given as
\begin{align}
\mathcal{L}_{CLFV}=&\mathcal{L}_\mathrm{photo}+\mathcal{L}_\mathrm{contact}, \label{eq:Lagrangian_clfv} \\
\mathcal{L}_\mathrm{photo}=&-\frac{4G_F}{\sqrt{2}}m_\mu
\left[A_R\overline{e_L}\sigma^{\mu\nu}\mu_R+A_L\overline{e_R}\sigma^{\mu\nu}\mu_L\right]F_{\mu\nu}+[\mathrm{H.c.}],
\label{eq:longrange}\\
\mathcal{L}_\mathrm{contact}=&-\frac{4G_F}{\sqrt{2}}
[g_1(\overline{e_L}\mu_R)(\overline{e_L}e_R)
+g_2(\overline{e_R}\mu_L)(\overline{e_R}e_L) \nonumber\\
&+g_3(\overline{e_R}\gamma_\mu\mu_R)(\overline{e_R}\gamma^\mu e_R)
+g_4(\overline{e_L}\gamma_\mu\mu_L)(\overline{e_L}\gamma^\mu e_L) \nonumber\\
&+g_5(\overline{e_R}\gamma_\mu\mu_R)(\overline{e_L}\gamma^\mu e_L)
+g_6(\overline{e_L}\gamma_\mu\mu_L)(\overline{e_R}\gamma^\mu e_R)]+[\mathrm{H.c.}],
\label{eq:shortrange}
\end{align}
where $G_F=1.166\times10^{-5}$GeV$^{-2}$ is the Fermi coupling constant,
and $A_{L/R}$ and $g_j$ ($j=1,\cdots,6$) are the dimensionless coupling constants.
The left- and right-handed projections are given as $P_{L/R}=(1\mp\gamma_5)/2$, respectively.
The $\mathcal{L}_\mathrm{contact}$ represents short range interaction and the $\mathcal{L}_\mathrm{photo}$
generates the one-photon-exchange process together with the ordinary electromagnetic interaction,
\begin{align}
\mathcal{L}_{em}=-q_e\overline{e}\gamma^\lambda eA_\lambda,
\end{align}
where $q_e=-e$ is a charge of an electron.

There are several clear differences on the interaction type.
For example, the atomic number ($Z$) dependence of the transition rate
is clearly different between photonic and contact interaction,
though it was shown that it is proportional to $(Z-1)^3$ in both cases
with the simple estimation given in Ref.~\cite{Koike2010}.
The cubic power of the atomic number arises from
the wave functions of the initial bound states. However, we 
showed in our previous works \cite{Uesaka2016,Uesaka2018} that the $Z$
dependence is stronger and weaker than $(Z-1)^3$ for the contact and
photonic interaction, respectively, by taking into account appropriate
wave functions and a photon propagator for photonic interaction.
As another example, in Ref.~\cite{Uesaka2018} it was
shown that the energy-angular distribution of emitted electrons is
sensitive to the interaction type.
With these facts, we can find which interaction is dominant.

However, the above observables do not depend on the chiral structure.
The chiral structure of the CLFV interaction is an important key to search
for the new physics.  For example, it is well known that SU(5) and SO(10)
supersymmetric grand unified theory gives different chiral structures in
the CLFV interaction.  To observe it in $\mu\to e\gamma$ and $\mu\to
eee$, to make use of the muon polarization has been discussed in
Refs.~\cite{Kuno1996,Okada2000}.  In this paper, we focus on the
\muetoee search with a muon polarized in a muonic atom to extract the
chiral property of the CLFV interaction.  We start from our previous
formulation of the \muetoee \cite{Uesaka2016,Uesaka2018}
and then we extended
the formalism to describe the electron asymmetry for the decay of a polarized
muon in an atom.
To determine the chiral structure of the CLFV interaction including parity violation, we investigate how the anisotropy in an electron momentum distribution depends on it.

In addition to the parity violating signal, we also analyze
motion-reversal-odd observables.  Even if the interaction Lagrangian
includes no $CP$ violating terms, the final state interaction and other
causes are known to induce the spurious $CP$ violation in observables.
Therefore the estimation of the contribution for the spurious $CP$
violation is essential to measure the $CP$ violation of the CLFV
interaction via the \muetoee process in the future.

In Sec.~\ref{sec:Formulation}, we formulate the \muetoee process in a muonic
atom with a polarized bound muon.  After we present the analytic formula
for the plane wave approximation to understand the mechanism of electron
asymmetry, we study the formula including the distortion of emitted
electrons.  In Sec.~\ref{sec:Results}, the results of
numerical calculation are given and the possibilities to identify the
structure of the CLFV operator are discussed.  Finally,
in Sec.~\ref{sec:Conclusion}, we summarize our analyses.

\section{Formulation \label{sec:Formulation}}

The \muetoee decay of the muonic atom is described in the independent
particle model of a muonic atom.  The transition amplitude $M$ of
$\mu^-(1s,s_\mu) + e^-(1s,s_e) \to e^-(\bm{p}_1,s_1) +
e^-(\bm{p}_2,s_2)$ is given as
\begin{align}
  2\pi i\delta(E_1+E_2-E_{tot})M(\bm{p}_1,s_1,\bm{p}_2,s_2;s_\mu,s_e)=&
  \braket{e^{s_1}_{\bm{p}_1}e^{s_2}_{\bm{p}_2}|T\left[\exp\left\{i\int d^4x(\mathcal{L}_{CLFV}+\mathcal{L}_{em})\right\}\right]|\mu^{s_\mu}_{1s}e^{s_e}_{1s}},
\end{align}
where we retain the first-order terms of the CLFV interaction.  $s_i$
($i=1,2$) is a spin of a scattering electron $i$ and $s_\ell$ ($\ell=e,\mu$) is a spin
of a bound lepton $\ell$. We define energies of emitted electrons
$E_i=p_i^0=\sqrt{p_i^2+m_e^2}$ and their maximum energy
$E_{tot}=m_\mu-B_\mu+m_e-B_e$, where $B_\ell$ is the binding energy of a
bound lepton $\ell$ in a $1s$ state.
Since a muon trapped by a nucleus
is rapidly deexcited into the ground state, it is sufficient to consider
only the case where the muon is in a $1s$ state.  In this paper, we only take
into account bound electrons in $1s$ states because they make the
dominant contribution for both photonic and contact processes.
The explicit form of transition matrix
$M$ is given in Refs.~\cite{Uesaka2016,Uesaka2018}.

The decay rate of a muonic atom with a polarized muon is given as
\begin{align}
\frac{d\Gamma}{d\epsilon_1d\Omega_1d\Omega_2}=&\frac{E_{tot}-2m_e}{128\pi^5}\left|\bm{p}_1\right|\left|\bm{p}_2\right| \nonumber\\
&\times\sum_{s_1,s_2}\sum_{s_e}\sum_{s_\mu,s_\mu'}M\left(\bm{p}_1,s_1,\bm{p}_2,s_2;s_\mu,s_e\right)\braket{s_\mu|\rho_\mu|s_\mu'}M^*\left(\bm{p}_1,s_1,\bm{p}_2,s_2;s_\mu',s_e\right),
\label{eq:differential_decay_rate_polarized}
\end{align}
where the muon spin density $\rho_\mu$ is represented by using
the muon polarization vector $\bm{P}$ as
\begin{align}
\rho_\mu=\frac{\bm{1}+\bm{\sigma}\cdot\bm{P}}{2}.
\label{eq:muon_spin_density}
\end{align}
Here $\bm{\sigma}=(\sigma_1,\sigma_2,\sigma_3)$ is the Pauli matrix.
We introduce dimensionless energies of electrons $\epsilon_i$ ($i=1,2$)
normalized by the maximum kinetic energy of final electrons as
\begin{align}
\epsilon_i=\frac{E_i-m_e}{E_{tot}-2m_e},
\end{align}
so that $0 \le \epsilon_i \le 1$ and $\epsilon_1+\epsilon_2=1$.
The differential decay rate for a polarized muon can be generally expressed by two functions $F(\epsilon_1,c_{12})$ and $F_D(\epsilon_1,c_{12})$ as
\begin{align}
  \frac{d\Gamma}{d\epsilon_1d\Omega_1d\Omega_2}=&
  \frac{1}{8\pi^2}\frac{d\Gamma_{unpol.}}{d\epsilon_1dc_{12}}
  \left[1+F\left(\epsilon_1,c_{12}\right)\bm{P}\cdot\hat{p}_1
    +F\left(\epsilon_2,c_{12}\right)\bm{P}\cdot\hat{p}_2
    +F_D\left(\epsilon_1,c_{12}\right)\bm{P}\cdot\left(\hat{p}_1\times\hat{p}_2\right)\right],
\label{eq:general_form_for_mu_pol}
\end{align}
where $\hat{p}_i$ is a unit vector in the direction of $\bm{p}_i$, and $c_{12}=\hat{p}_1\cdot\hat{p}_2$ is the cosine of an angle between emitted electrons.
$\Gamma_{unpol.}$ is the rate of \muetoee for an unpolarized muon, which is given in our previous works \cite{Uesaka2016,Uesaka2018} (see also the Appendix in
this paper).
Since the differential decay rate must be symmetric under
the exchange of $\bm{p}_1$ and $\bm{p}_2$, the coefficients of
the $\bm{P}\cdot\hat{p}_1$ and $\bm{P}\cdot\hat{p}_2$ terms are written by
the same function $F$.
The coefficient $F_D$ of the $\bm{P}\cdot\left(\hat{p}_1\times\hat{p}_2\right)$ term must satisfy
\begin{align}
F_D\left(\epsilon_2,c_{12}\right)=-F_D\left(\epsilon_1,c_{12}\right).
\label{eq:symmetry_for_tildef}
\end{align}
From Eq.~(\ref{eq:general_form_for_mu_pol}), it is found that the effect of muon polarization disappears when
$\bm{p}_1+\bm{p}_2=0$.

\subsection{Plane wave approximation \label{subsec:pwa}}

Firstly, we examine the transition amplitude for contact interaction
with a plane wave approximation for the final scattered electrons and
examine an analytic form of $F$'s to understand the origin of asymmetry.
In this section, we derive asymmetry with a single operator dominance
hypothesis and also neglect the masses of the final electrons.  The transition
matrix element of contact interaction with scalar coupling
is written
by using helicity representation $h_i$ of scattered electrons as
\begin{align}
  M(\bm{p}_1,h_1,\bm{p}_2,h_2;s_\mu,s_e)
 =& - \frac{G_F}{\sqrt{2}} g
 \int d^3r\left[
 \left(\chi_{\hat{p}_1}^{h_1\dagger},-h_1\chi_{\hat{p}_1}^{h_1\dagger}\right)
 \left(1-h_a\gamma_5\right)\psi_\mu^{s_\mu}(\bm{r})\right] \nonumber\\
 &\times\left[
 \left(\chi_{\hat{p}_2}^{h_2\dagger},-h_2\chi_{\hat{p}_2}^{h_2\dagger}\right)
 \left(1-h_b\gamma_5\right)\psi_e^{s_e}(\bm{r})\right]
 e^{-i \bm{p}\cdot\bm{r}}
  -\left(\left\{p_1,h_1\right\}\leftrightarrow \left\{p_2,h_2\right\}\right),
\label{eq:scalar_contact}
\end{align}
where $\bm{p}=\bm{p}_1+\bm{p}_2$.
Chirality of the interaction determines the constants
$h_a$ and $h_b$. For example,
\begin{align}
\{h_a,h_b,g\}=
\begin{cases}
\{-1,-1,g_1\} & \text{for } g_1\ne 0, g_{j\ne 1}=0 \\
\{+1,+1,g_2\} & \text{for } g_2\ne 0, g_{j\ne 2}=0
\end{cases}.
\end{align}
The wave functions of a bound lepton are written as
\begin{align}
\psi_\ell^s(\bm{r})=&
\begin{pmatrix}
g_\ell(r)\chi^s \\
-if_\ell(r)\sigma\cdot\hat{r}\chi^s
\end{pmatrix},
\end{align}
where $\chi^s$ is a two-component spinor.
The transition matrix, Eq.~(\ref{eq:scalar_contact}), can be expressed as
\begin{align}
  M(\bm{p}_1,h_1,\bm{p}_2,h_2;s_\mu,s_e)
  =
  - \frac{G_F}{\sqrt{2}}g
  \delta_{h_a,h_1}\delta_{h_b,h_2}&
  \left[\left\{(\chi^{h_1\dagger}_{\hat{p}_1}\chi^{s_\mu})I_{gg}+h_a(\chi^{h_1\dagger}_{\hat{p}_1}\sigma\cdot\hat{p}\chi^{s_\mu})I_{fg}\right\}
    (\chi^{h_2\dagger}_{\hat{p}_2}\chi^{s_e}\right.) \nonumber\\
     &+\left\{(\chi^{h_1\dagger}_{\hat{p}_1}\chi^{s_\mu})I_{gf}+h_a(\chi^{h_1\dagger}_{\hat{p}_1}\sigma\cdot\hat{p}\chi^{s_\mu})I_{ff}\right\}
    h_b(\chi^{h_2\dagger}_{\hat{p}_2}\sigma\cdot\hat{p}\chi^{s_e}) \nonumber\\
    &\left. - h_ah_b(\chi^{h_1\dagger}_{\hat{p}_1}\sigma\chi^{s_\mu})\cdot(\chi^{h_2\dagger}_{\hat{p}_2}\sigma\chi^{s_e})\tilde{I}_{ff}\right]-\left(\left\{p_1,h_1\right\}\leftrightarrow\left\{p_2,h_2\right\}\right).
  \label{eq:me-plw}
\end{align}
Here we define the radial integrals as
\begin{align}
I_{gg}=& 4\pi\int drr^2g_\mu(r)g_e(r)j_0(pr), \\
I_{fg}=& 4\pi\int drr^2f_\mu(r)g_e(r)j_1(pr), \\
I_{gf}=& 4\pi\int drr^2g_\mu(r)f_e(r)j_1(pr), \\
I_{ff}=& 4\pi\int drr^2f_\mu(r)f_e(r)j_2(pr), \\
\tilde{I}_{ff}=& 4\pi\int drr^2f_\mu(r)f_e(r)\frac{j_1(pr)}{pr},
\end{align}
where $j_n$ is the $n$th-order spherical Bessel function.
It is straightforward to evaluate asymmetry functions
from the transition matrix element in Eq. (\ref{eq:me-plw}), which is used
to test the complicated multipole expansion formula
in the next subsection.

The integrals $I_{gg}$, $I_{fg}$, and $I_{gf}$ are comparable in magnitude, while $I_{ff}\sim\tilde{I}_{ff}\sim Z\alpha I_{gg}$.
By neglecting $I_{ff}$ and $\tilde{I}_{ff}$,
the asymmetry function Eq.~(\ref{eq:general_form_for_mu_pol})
for a $g_j$-type ($j=1,2$) interaction is given as 
\begin{align}
  F(\epsilon_1,c_{12}) \simeq &\frac{h_a\epsilon_1}{d}\frac{2I_{gg}\left(I_{fg}-I_{gf}\right)}{I_{gg}^2+\left(I_{fg}-I_{gf}\right)^2}.
  \label{eq:F12g}
\end{align}
Here
\begin{align}
d=\sqrt{1-2\epsilon_1\epsilon_2\left(1-c_{12}\right)}.
\end{align}
The vector-type interactions $g_3$ and $g_4$ give very similar
results as $g_1$ and $g_2$ interactions, though the exact analytic formula
is slightly different from Eq. (\ref{eq:F12g}).
An important finding is that $F$ is proportional to $h_a$.
That is, the asymmetry reveals the chiral structure of CLFV interaction.
This holds even for exact (distorted) wave functions of electrons.
It is also noted that
$I_{fg}\simeq I_{gf}$
for bound state lepton wave functions with point nuclear charge density.
Therefore the use of a finite nuclear charge distribution,
which makes a difference between the muon and electron bound state wave function,
is essential to obtain measurable asymmetry for $g_1$-$g_4$ interactions.

The vector-type interactions $g_5$ and $g_6$ in Eq.~(\ref{eq:shortrange}) can also be written as scalar interactions using the Fierz transformation:
\begin{align}
g_5\left(e_R\gamma_\mu\mu_R\right)\left(e_L\gamma^\mu e_L\right)=-2g_5\left(e_L\mu_R\right)\left(e_Re_L\right), \\
g_6\left(e_L\gamma_\mu\mu_L\right)\left(e_R\gamma^\mu e_R\right)=-2g_6\left(e_R\mu_L\right)\left(e_Le_R\right).
\end{align}
With the relation, we can directly apply Eqs. (\ref{eq:scalar_contact}) and (\ref{eq:me-plw}) by assigning
\begin{align}
\{h_a,h_b,g\}=
\begin{cases}
\{-1,+1,-2g_5\} & \text{for } g_5\ne 0, g_{j\ne 5}=0 \\
\{+1,-1,-2g_6\} & \text{for } g_6\ne 0, g_{j\ne 6}=0
\end{cases}.
\end{align}
Again, by neglecting the higher-order contribution of $Z\alpha$, we obtain
asymmetry for the $g_5$- and $g_6$-type interactions as
\begin{align}
F(\epsilon_1,c_{12}) \simeq &\frac{h_a}{D}\left[\frac{I_{gg}^2-I_{fg}^2+I_{gf}^2}{2}+2I_{gg}I_{fg}\frac{\epsilon_1}{d}+I_{gg}I_{gf}\frac{\epsilon_2+\epsilon_1c_{12}}{d}+I_{fg}\left(I_{fg}+I_{gf}\right)\frac{\epsilon_1\left(1+c_{12}\right)}{d^2}\right], \\
D=& I_{gg}^2+I_{fg}^2+I_{gf}^2+I_{gg}\left(I_{fg}+I_{gf}\right)\frac{1+c_{12}}{d}+2I_{fg}I_{gf}\frac{\left(\epsilon_1+\epsilon_2c_{12}\right)\left(\epsilon_2+\epsilon_1c_{12}\right)}{d^2}.
\label{eq:F56g}
\end{align}
The asymmetry is again proportional to $h_a$, but it
remains finite even for wave functions under point charge.

\subsection{Multipole expansion}

Following our previous works in Refs.~\cite{Uesaka2016,Uesaka2018},
we introduce partial wave expansion of the scattering electron states
and the transition amplitude $M$ in Eq. (\ref{eq:differential_decay_rate_polarized}) is written 
as
\begin{align}
  M(\bm{p}_1,s_1,\bm{p}_2,s_2;s_\mu,s_e)=& 2\sqrt{2}G_F
  \sum_{\kappa_1,\kappa_2,\nu_1,\nu_2,m_1,m_2}
  \left(4\pi\right)^2\left(-i\right)^{l_{\kappa_1}+l_{\kappa_2}}
  e^{i\left(\delta_{\kappa_1}+\delta_{\kappa_2}\right)} \nonumber\\
  &\times Y_{l_{\kappa_1}}^{m_1}\left(\hat{p}_1\right)Y_{l_{\kappa_2}}^{m_2}
  \left(\hat{p}_2\right)\left(l_{\kappa_1},m_1,1/2,s_1|j_{\kappa_1},\nu_1\right)
  \left(l_{\kappa_2},m_2,1/2,s_2|j_{\kappa_2},\nu_2\right) \nonumber\\
&\times\sum_{J,M}\left(j_{\kappa_1},\nu_1,j_{\kappa_2},\nu_2|J,M\right)\left(j_{-1},s_\mu,j_{\kappa_e},s_e|J,M\right) \nonumber\\
&\times\frac{\sqrt{2\left(2j_{\kappa_1}+1\right)\left(2j_{\kappa_2}+1\right)\left(2j_{\kappa_e}+1\right)}}{4\pi}N\left(J,\kappa_1,\kappa_2,E_1,\alpha_e\right),
\label{eq:M_photo}
\end{align}
where $\left(l_\kappa,m,1/2,s|j_\kappa,\nu\right)$ and $Y_{l_\kappa}^m(\hat{p})$ are the Clebsch-Gordan coefficients and the spherical harmonics, respectively.
$N\left(J,\kappa_1,\kappa_2,E_1,\alpha_e\right)$ is defined in Ref.~\cite{Uesaka2018}, which is shown in the Appendix.
After straightforward calculation, it is found that
the asymmetry functions $F$ and $F_D$, defined in Eq.~(\ref{eq:general_form_for_mu_pol}), can be written as follows:
\begin{align}
F\left(\epsilon_1,c_{12}\right)=&\frac{\displaystyle\sum_{l\ge 1}c_l^F\left(\epsilon_1\right)P_{l}'\left(c_{12}\right)}{\displaystyle\sum_{l\ge 0}c_l\left(\epsilon_1\right)P_{l}\left(c_{12}\right)},
\label{eq:F}
\end{align}
and
\begin{align}
F_D\left(\epsilon_1,c_{12}\right)=&\frac{\displaystyle\sum_{l\ge 1}c_l^{F_D}\left(\epsilon_1\right)P_{l}'\left(c_{12}\right)}{\displaystyle\sum_{l\ge 0}c_l\left(\epsilon_1\right)P_{l}\left(c_{12}\right)},
\label{eq:F_tilde}
\end{align}
where $P_l$ and $P_l'$ are the Legendre polynomial and its derivative, respectively.
The coefficients are given as
\begin{align}
c_l^F\left(\epsilon_1\right)=&\left(E_{tot}-2m_e\right)\frac{G_F^2}{2\pi^3}\left|\bm{p}_1\right|\left|\bm{p}_2\right|\left(2j_{\kappa_e}+1\right) \nonumber\\
&\times\sqrt{6}\sum_{\kappa_1,\kappa_2}\sum_{\kappa_1',\kappa_2'}\sum_{J,J'}(-1)^{J'+j_{\kappa_1'}-j_{\kappa_2'}-j_{\kappa_e}+l+1/2}\frac{1+(-1)^{l_{\kappa_1}+l_{\kappa_1'}+l}}{2}\frac{1-(-1)^{l_{\kappa_2}+l_{\kappa_2'}+l}}{2} \nonumber\\
&\times i^{-l_{\kappa_1}-l_{\kappa_2}+l_{\kappa_1'}+l_{\kappa_2'}}e^{i\left(\delta_{\kappa_1}+\delta_{\kappa_2}-\delta_{\kappa_1'}-\delta_{\kappa_2'}\right)}N(J,\kappa_1,\kappa_2,E_1,\alpha_e)N^*(J',\kappa'_1,\kappa'_2,E_1,\alpha_e) \nonumber\\
&\times(2J+1)(2J'+1)\left(2j_{\kappa_1}+1\right)\left(2j_{\kappa_2}+1\right)\left(2j_{\kappa_1'}+1\right)\left(2j_{\kappa_2'}+1\right) \nonumber\\
&\times
\begin{Bmatrix}
J & J' & 1 \\
1/2 & 1/2 & j_{\kappa_e}
\end{Bmatrix}
\sqrt{2l+1}
\begin{pmatrix}
j_{\kappa_1} & j_{\kappa'_1} & l \\
1/2 & -1/2 & 0
\end{pmatrix} \nonumber\\
&\times\left[\sqrt{\frac{2l+3}{l+1}}
\begin{pmatrix}
j_{\kappa_2} & j_{\kappa'_2} & l+1 \\
1/2 & -1/2 & 0
\end{pmatrix}
\begin{Bmatrix}
j_{\kappa_1} & j_{\kappa_2} & J \\
j_{\kappa_1'} & j_{\kappa_2'} & J' \\
l & l+1 & 1
\end{Bmatrix}\right. \nonumber\\
&\left.+\sqrt{\frac{2l-1}{l}}
\begin{pmatrix}
j_{\kappa_2} & j_{\kappa'_2} & l-1 \\
1/2 & -1/2 & 0
\end{pmatrix}
\begin{Bmatrix}
j_{\kappa_1} & j_{\kappa_2} & J \\
j_{\kappa_1'} & j_{\kappa_2'} & J' \\
l & l-1 & 1
\end{Bmatrix}\right],
\end{align}
and
\begin{align}
c_l^{F_D}\left(\epsilon_1\right)=&\left(E_{tot}-2m_e\right)\frac{G_F^2}{2\pi^3}\left|\bm{p}_1\right|\left|\bm{p}_2\right|\left(2j_{\kappa_e}+1\right) \nonumber\\
&\times\sqrt{6}\sum_{\kappa_1,\kappa_2}\sum_{\kappa_1',\kappa_2'}\sum_{J,J'}(-1)^{J'+j_{\kappa_1'}-j_{\kappa_2'}-j_{\kappa_e}+l-1/2}\frac{1+(-1)^{l_{\kappa_1}+l_{\kappa_1'}+l}}{2}\frac{1+(-1)^{l_{\kappa_2}+l_{\kappa_2'}+l}}{2} \nonumber\\
&\times i^{1-l_{\kappa_1}-l_{\kappa_2}+l_{\kappa_1'}+l_{\kappa_2'}}e^{i\left(\delta_{\kappa_1}+\delta_{\kappa_2}-\delta_{\kappa_1'}-\delta_{\kappa_2'}\right)}N(J,\kappa_1,\kappa_2,E_1,\alpha_e)N^*(J',\kappa'_1,\kappa'_2,E_1,\alpha_e) \nonumber\\
&\times(2J+1)(2J'+1)\left(2j_{\kappa_1}+1\right)\left(2j_{\kappa_2}+1\right)\left(2j_{\kappa_1'}+1\right)\left(2j_{\kappa_2'}+1\right) \nonumber\\
&\times
\begin{Bmatrix}
J & J' & 1 \\
1/2 & 1/2 & j_{\kappa_e}
\end{Bmatrix}
\frac{(2l+1)^{3/2}}{\sqrt{l(l+1)}}
\begin{pmatrix}
j_{\kappa_1} & j_{\kappa'_1} & l \\
1/2 & -1/2 & 0
\end{pmatrix}
\begin{pmatrix}
j_{\kappa_2} & j_{\kappa'_2} & l \\
1/2 & -1/2 & 0
\end{pmatrix}
\begin{Bmatrix}
j_{\kappa_1} & j_{\kappa_2} & J \\
j_{\kappa_1'} & j_{\kappa_2'} & J' \\
l & l & 1
\end{Bmatrix},
\end{align}
where the $3j$, $6j$, and $9j$ symbols are used.
Here $\alpha_e$ and $\kappa_e$ indicate
the state and the angular momentum of the bound electron, respectively.
Although $\alpha_e=1s$ and $\kappa_e=-1$ are assumed in this analysis,
those formulas can be applied to bound electrons in higher orbit.  The
common denominator of Eqs.~(\ref{eq:F}) and (\ref{eq:F_tilde}) gives the
differential decay rate for an unpolarized muon, whose explicit formula is
also given in the Appendix.

\section{Results and Discussions \label{sec:Results}}

\begin{figure}[htb]
  \centering
    \begin{tabular}{c}
      \begin{minipage}{0.45\hsize}
		\centering
          \includegraphics[clip, width=5.0cm]{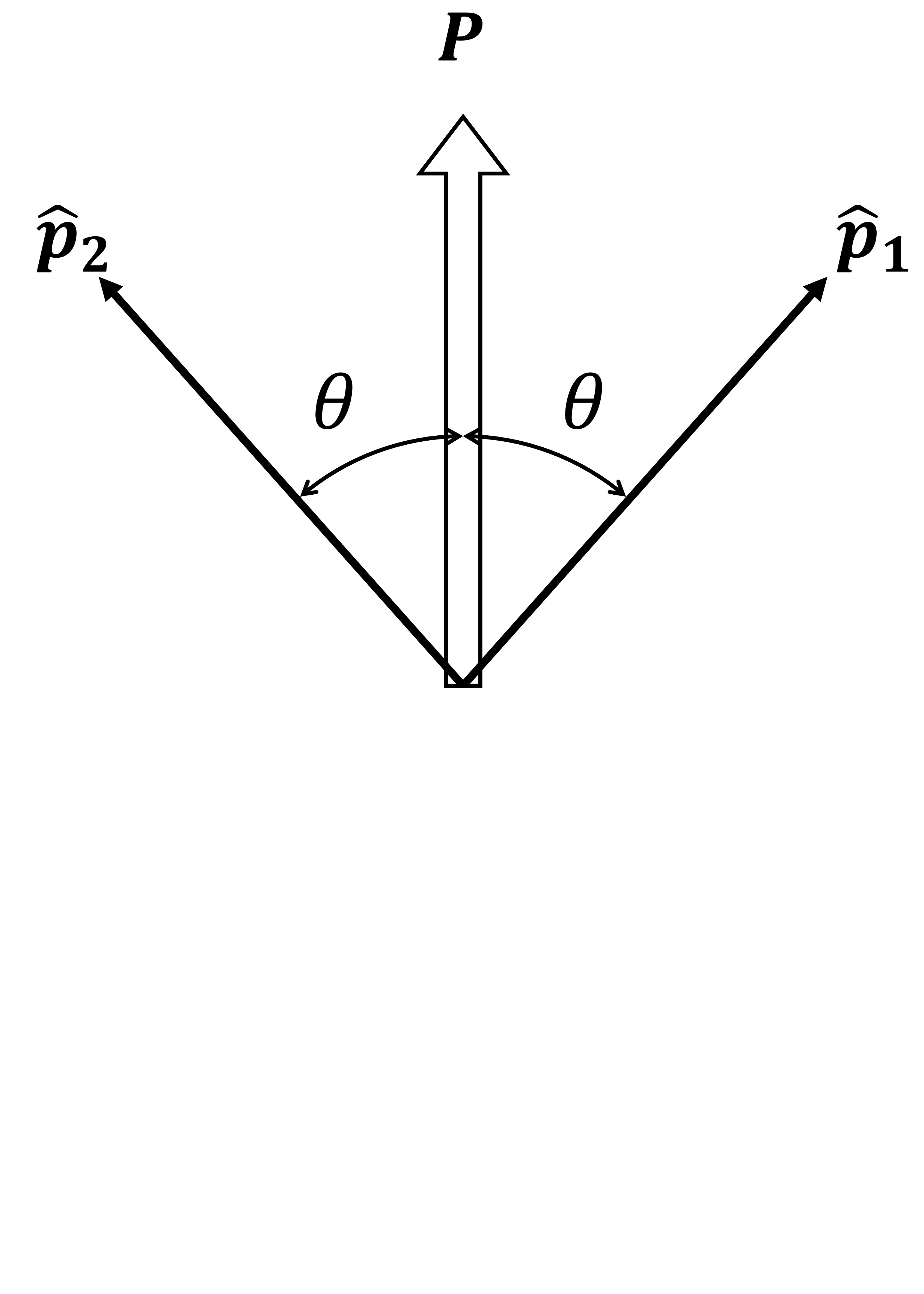}
          \\ (a)
      \end{minipage}%
      \begin{minipage}{0.45\hsize}
        \centering
          \includegraphics[clip, width=5.0cm]{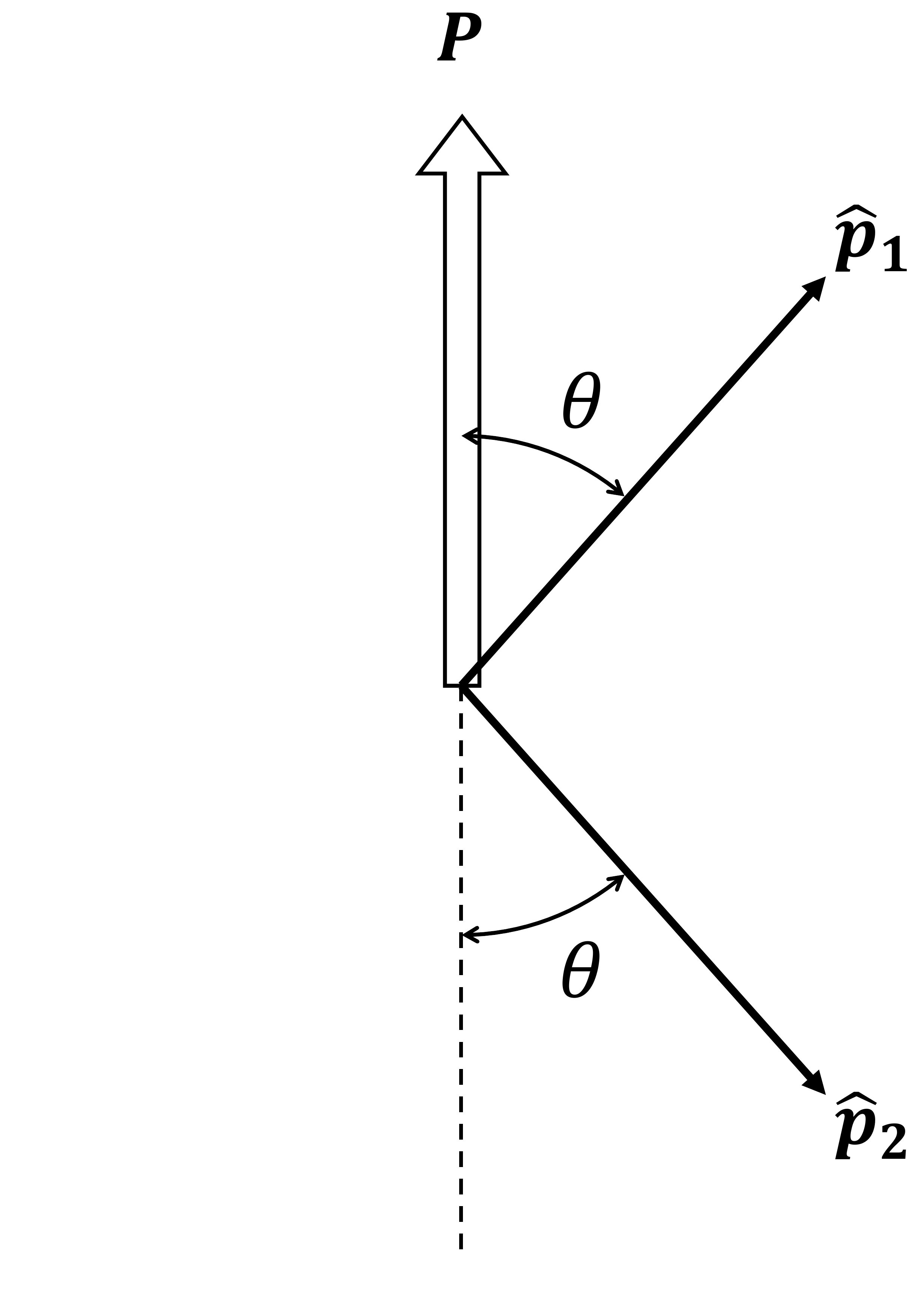}
          \\ (b)
      \end{minipage}%
    \end{tabular}
    \caption{Pictorial representation of the directions of muon polarization vector $\bm{P}$ and momentum of emitted electrons $\hat{p}_i$.
	(a) and (b) show the configurations where only $F_S$ and $F_A$ contribute, respectively.
	Here, three vectors, $\bm{P}$, $\hat{p}_1$, and $\hat{p}_2$, are in the same plane.
	$\theta$ is defined as an angle between $\bm{P}$ and $\hat{p}_1$.
	In (a), the vector $\hat{p}_1+\hat{p}_2$ is parallel to $\bm{P}$ so that the angle between $\hat{p}_1$ and $\hat{p}_2$ is $2\theta$.
	In (b), $\hat{p}_1+\hat{p}_2$ is perpendicular to $\bm{P}$ so that the angle between $\hat{p}_1$ and $\hat{p}_2$ is $\pi-2\theta$.}
    \label{fig:vector}
\end{figure}

The angular distribution of electrons from a polarized muon can be decomposed into three independent components as
\begin{align}
  \frac{d\Gamma}{d\epsilon_1d\Omega_1d\Omega_2} =
  \frac{d\Gamma_{unpol.}}{d\epsilon_1d\Omega_1d\Omega_2}
  \left[1+F_S\left(\epsilon_1,c_{12}\right)\bm{P}\cdot\hat{p}_{12}+F_A\left(\epsilon_1,c_{12}\right)\bm{P}\cdot\hat{q}_{12}+F_D\left(\epsilon_1,c_{12}\right)\bm{P}\cdot\left(\hat{p}_1\times\hat{p}_2\right)\right].
\end{align}
Here, $\hat{p}_{12}=\left(\hat{p}_1+\hat{p}_2\right)/\left|\hat{p}_1+\hat{p}_2\right|$ and $\hat{q}_{12}=\left(\hat{p}_1-\hat{p}_2\right)/\left|\hat{p}_1-\hat{p}_2\right|$, and
\begin{align}
F_S\left(\epsilon_1,c_{12}\right)=&\sqrt{\frac{1+c_{12}}{2}}\left[F\left(\epsilon_1,c_{12}\right)+F\left(\epsilon_2,c_{12}\right)\right], \\
F_A\left(\epsilon_1,c_{12}\right)=&\sqrt{\frac{1-c_{12}}{2}}\left[F\left(\epsilon_1,c_{12}\right)-F\left(\epsilon_2,c_{12}\right)\right].
\end{align}
$F_D$ shows the asymmetry when the polarization of muon
 $\bm{P}$ is perpendicular to the momentum of emitted electrons.
$F_S$ and $F_A$ show the asymmetry
when $\bm{P}$ is in the plane of the momentum of emitted electrons.
$F_S$ ($F_A$) represents the electron asymmetry when $\hat{p}_1+\hat{p}_2$ is parallel (perpendicular)
to the polarization vector $\bm{P}$ and $c_{12}=\cos 2\theta$($c_{12}=-\cos 2\theta$),
where $\theta$ is defined as an angle between $\bm{P}$ and $\hat{p}_1$
as shown in Fig.~\ref{fig:vector}.

We consider the following cases in the single operator dominance hypothesis:
\begin{enumerate}
\item Contact interaction with scalar coupling, where the electrons are emitted with the same chirality:
\begin{align}
g_1 = 1, \hspace{1mm} A_{L/R}=0, \text{ and } g_{j\ne 1}=0.
\label{eq:case1}
\end{align}
\item Contact interaction with vector coupling, where the electrons are emitted with the same chirality:
\begin{align}
g_3= 1, \hspace{1mm} A_{L/R}=0, \text{ and } g_{j\ne 1}=0.
\label{eq:case2}
\end{align}
\item Contact interaction, where the electrons are emitted with the opposite chirality:
\begin{align}
g_5 = 1, \hspace{1mm} A_{L/R}=0, \text{ and } g_{j\ne 5}=0.
\label{eq:case3}
\end{align}
\item Photonic interaction:
\begin{align}
A_R = 1, \hspace{1mm} A_L=0, \text{ and } g_{j}=0.
\label{eq:case4}
\end{align}
\end{enumerate}

In the following, we show results of asymmetry coefficients for
a polarized muon in $^{208}$Pb.
The muon and electron wave functions are obtained by solving a Dirac
equation with a Coulomb potential of the uniform distribution of
nuclear charge $\rho_C(r)$, 
\begin{align}
\rho_C(r)=\frac{3Ze}{4\pi R^3}\theta(R-r),
\end{align}
with $R=1.2A^{1/3}$fm, where $A$ is the mass number of the nucleus.
The numerical results using a multipole expansion formula
have been tested by comparing the results of a simple formula with the plane wave
approximation.

\begin{figure}[htb]
  \centering
\includegraphics[width=0.95\textwidth]{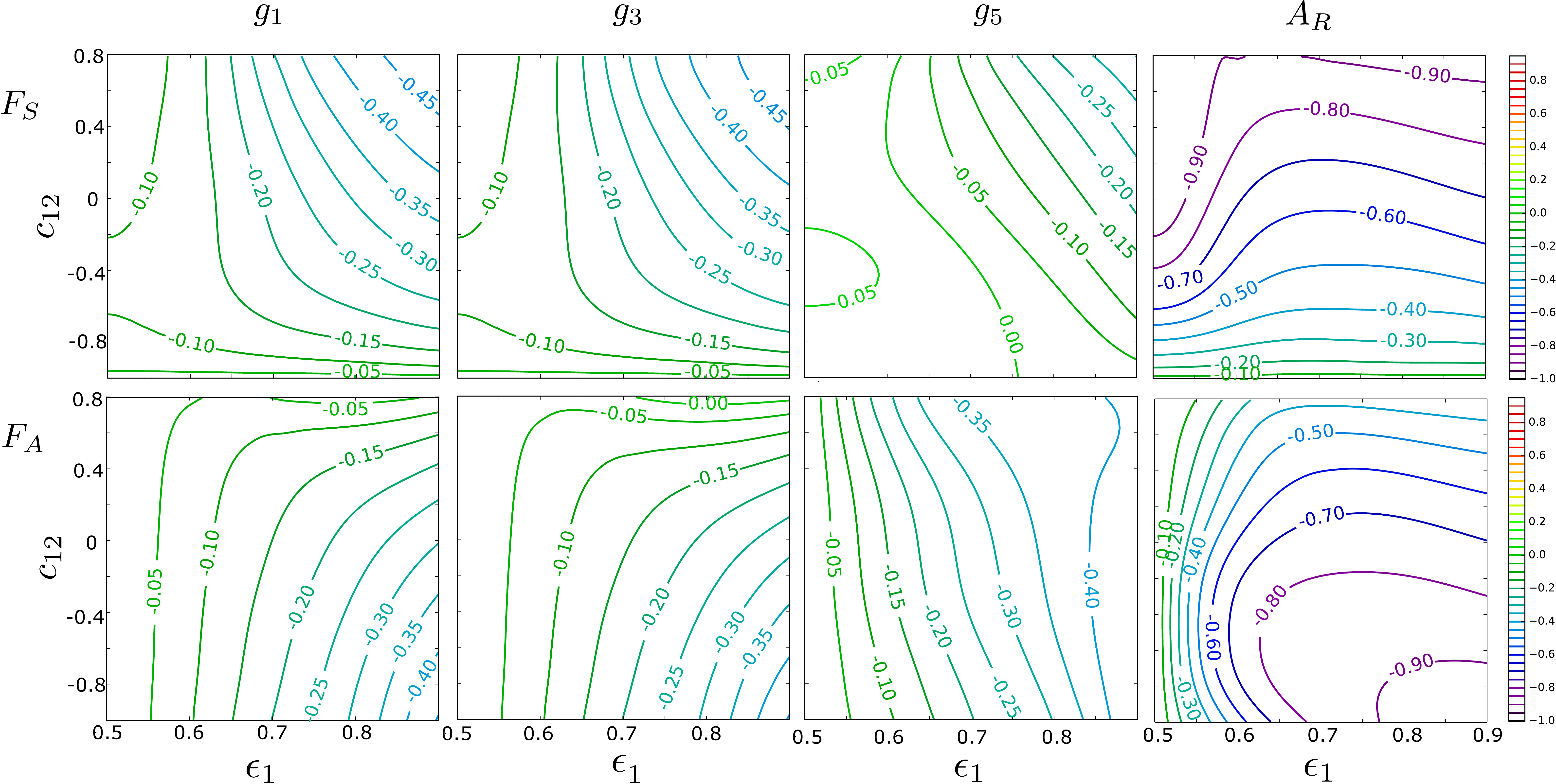}
\caption{The asymmetry functions $F_S$(upper panel) and $F_A$(lower panel)
  for $g_1$, $g_3$, $g_5$, and $A_R$.}
    \label{fig:FsFa}
\end{figure}

\begin{figure}[htb]
  \centering
\includegraphics[width=0.7\textwidth]{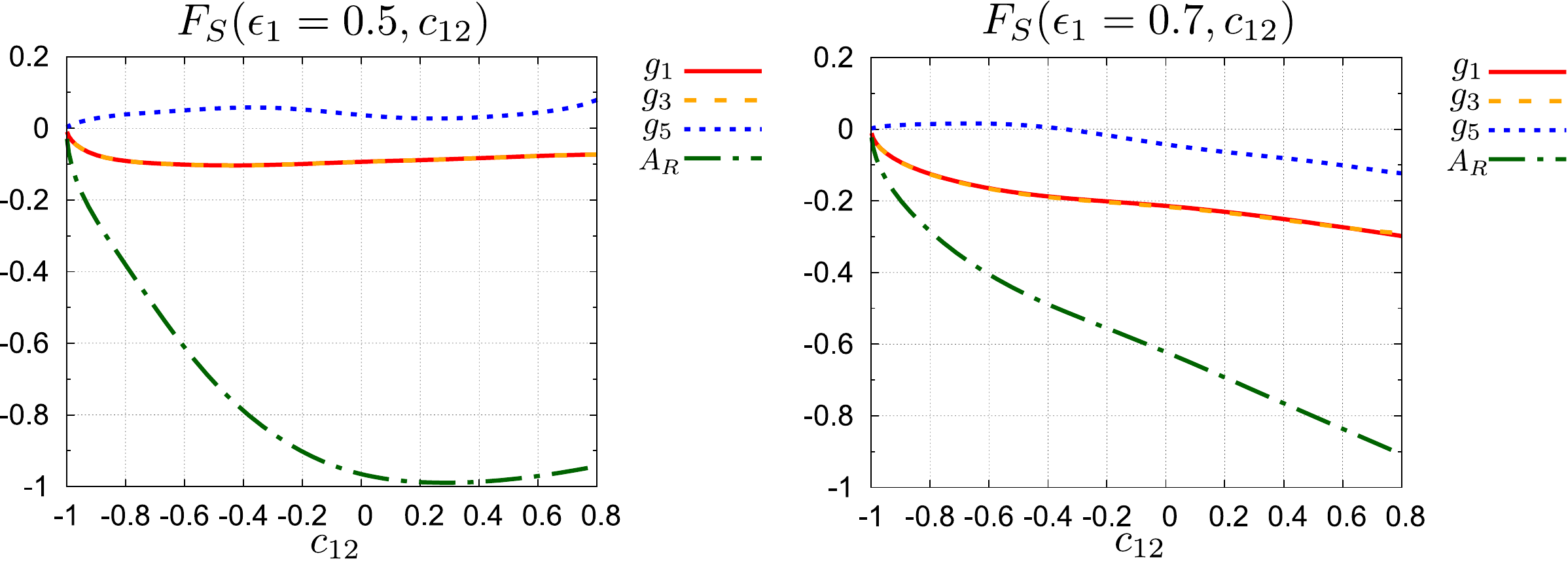}
\caption{The asymmetry function $F_S$ for $g_1$, $g_3$, $g_5$, and $A_R$, where $\epsilon_1$ is fixed at $0.5$ (left panel) and $0.7$ (right panel).}
    \label{fig:Fs2}
\end{figure}

\begin{figure}[htb]
  \centering
\includegraphics[width=0.7\textwidth]{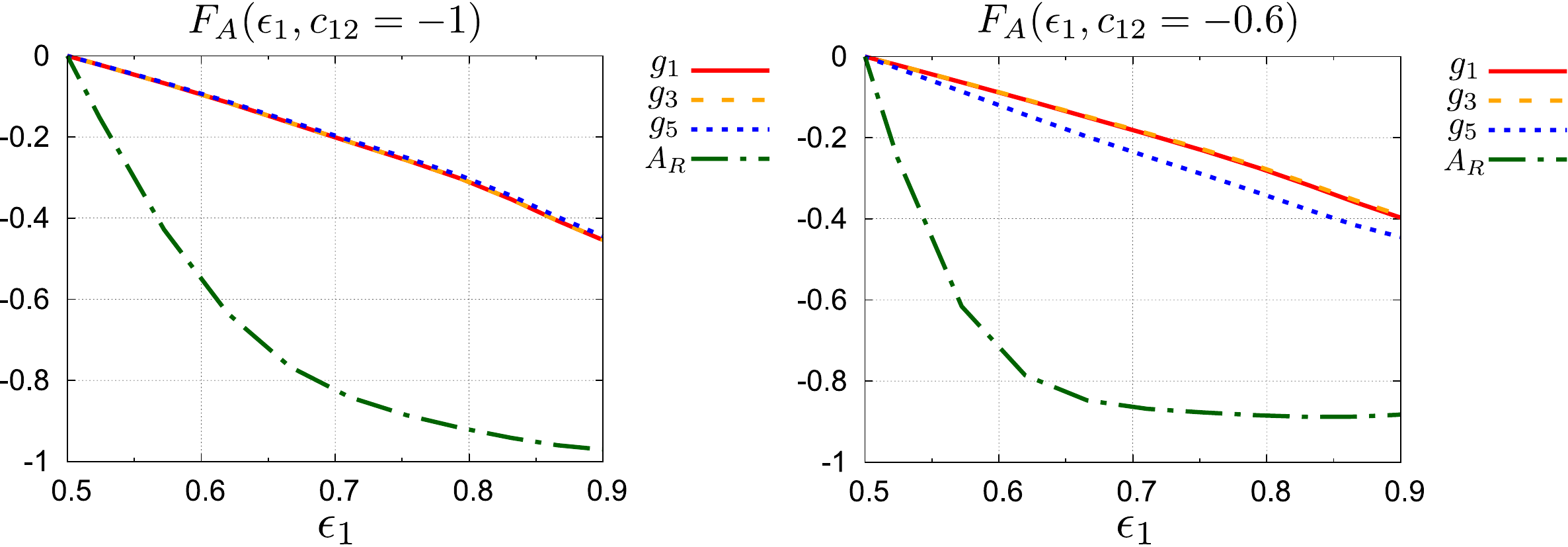}
\caption{The asymmetry function $F_A$ for $g_1$, $g_3$, $g_5$, and $A_R$, where $c_{12}$ is fixed at $-1$ (left panel) and $-0.6$ (right panel).}
    \label{fig:Fa2}
\end{figure}

The 2D plots of $F_S$ and $F_A$ for $g_1$, $g_3$, $g_5$, and $A_R$
as a function of electron energy $\epsilon_1$ and angle $c_{12}$
are shown in Fig.~\ref{fig:FsFa}.
It is sufficient to consider only the domain $0.5\le\epsilon_1\le 1$ for each function because, without loss of generality, the electron indices $i=1,2$ can be assigned so that $\epsilon_1\ge\epsilon_2$.
The values of the asymmetry functions of $g_2$, $g_4$, $g_6$, and $A_L$ are just
the opposite sign of those of $g_1$, $g_3$, $g_5$, and $A_R$ as discussed in the plane wave approximation.
Figure~\ref{fig:Fs2} shows the $c_{12}$ dependence of $F_{S}$ for fixed $\epsilon_1$,
 and Fig.~\ref{fig:Fa2} shows the $\epsilon_1$-dependence of $F_{A}$ for fixed $c_{12}$.
Depending on the CLFV interaction,
the asymmetry functions can have a large value in a certain 
domain of $\epsilon_1$ and $c_{12}$. The asymmetry of
photonic interaction $A_R$ is very different from those of contact
interactions. Furthermore,
the $g_1$ interaction can be distinguished from $g_5$
by using the $\epsilon_1$-$c_{12}$ dependence of $F_S$, while
the asymmetry functions of $g_1$ are almost the same as those of $g_3$.


The $F_D \bm{P}\cdot\left(\hat{p}_1\times\hat{p}_2\right)$ term 
 is parity even but motion-reversal odd \cite{time_reversal}.
 Nonzero $F_D$ could be used as a signal of the $CP$ violation of the CLFV interaction.
 However, it is known, for example in beta decay \cite{Ando2009}, that
 the final state interaction generates a motion-reversal-odd correlation.
 For the \muetoee process, the distortion effect of the emitted electrons by the nuclear potential also contributes to $F_D$.
In order to quantitatively determine the $CP$ violating phases of CLFV interactions, 
it is necessary to evaluate the asymmetry stemming from effects other than the $CP$ violation and to correctly count them as background contributions.

\begin{figure}[htb]
  \centering
\includegraphics[width=0.95\textwidth]{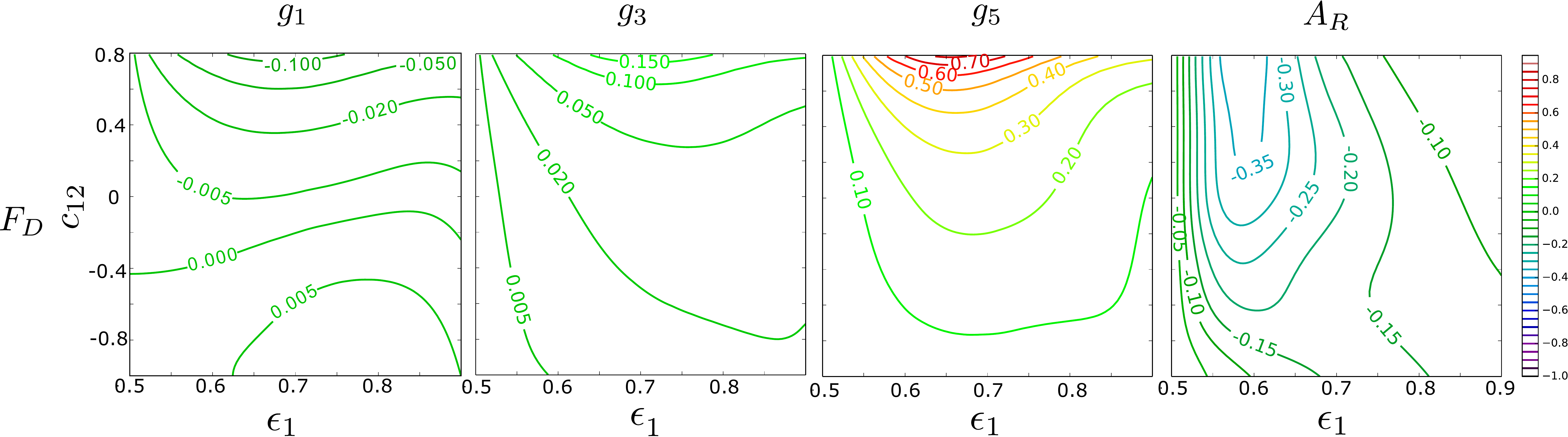}
    \caption{The asymmetry function $F_D$ for $g_1$, $g_3$, $g_5$, and $A_R$.}
    \label{fig:Fd}
\end{figure}

Although coupling constants ($g_j,A_{L/R}$) are taken to be real,
the coefficient $F_D$ of the $\bm{P}\cdot\left(\hat{p}_1\times\hat{p}_2\right)$ term
in Eq.~(\ref{eq:general_form_for_mu_pol}) becomes nonzero due to the final state interaction.
We concentrate on the final state interaction between each electron and nucleus, which has a much more significant effect than that between electrons because of the large nuclear charge.
In the photonic interaction, an additional source of nonzero $F_D$ is
the propagation of real and virtual photons in the intermediate state.
Figure~\ref{fig:Fd} shows the function $F_D$ as a function of $\epsilon_1$ and $c_{12}$.
Since $\bm{P}\cdot\left(\hat{p}_1\times\hat{p}_2\right)$ is even under parity transformation,
$F_D$ of $g_1$ is the same as that of $g_2$ in contrast to $F_S$ and $F_A$.
$F_D$ is small for $g_1$ and $g_3$ and it has the largest value in photonic interactions.
It is important to include the final state interaction properly in future searches
for $CP$ violation of the CLFV interaction.

\section{Conclusion \label{sec:Conclusion}}

We have studied the energy-angular distributions of emitted electrons of \muetoee decay in a muonic atom with a polarized muon.
The asymmetric distribution of electrons provides information about the CLFV interaction
in addition to the atomic number dependence of
the CLFV decay rate discussed in our previous works \cite{Uesaka2016,Uesaka2018}.
Our findings are as follows:
(i) The angular distribution of emitted electrons is a parity violating quantity.
Therefore the asymmetry functions $F_S$ and $F_A$ probe the parity violation
of CLFV interactions. 
(ii) The sign of the asymmetry directly reflects the chirality of the muon involved
in the \muetoee process.
(iii) Furthermore the $\epsilon_1$ and $c_{12}$ dependence of the asymmetry depends on the type of 
the CLFV operator in the effective Lagrangian, which would also be useful to distinguish
models beyond the SM. In addition,
it was found that the asymmetry could be large for the \muetoee process
induced by the photonic dipole interaction.

We have also estimated the asymmetry coefficient of the motion-reversal-odd term assuming no $CP$ violating CLFV interactions.
The asymmetry coefficient due to the final state interaction is not large, typically $O\left(10^{-2}\sim10^{-1}\right)$.

In order to measure the asymmetry with enough precision we need a sufficient number of events.
The \muetoee event rate is maximum at $\epsilon_1=0.5$ and $c_{12}=-1$ \cite{Uesaka2016,Uesaka2018}, where the asymmetry is zero.
Therefore more careful study is needed to find an optimal kinematical condition to realize the idea and efforts are in progress.

\begin{acknowledgments}

This work was supported by the JSPS KAKENHI Grants No.~18H01210 and No.~JP18H05543 (J.S.);
Grants No.~18H05543a, No.~16K05354, and No.~19H05104(T.S.); Grant No.~18H05231 (Y.K.);
and Grant No.~16K17693 (M.Y.), and the Sasakawa Scientific Research Grant from the Japan Science Society (Y.U.).

\end{acknowledgments}

\appendix
\section{COMPLETE FORMULAS FOR DECAY RATE \label{app:formulae}}

In this appendix, we show the formulas for the differential decay rate where the initial muon is unpolarized, which have been already given in Refs.~\cite{Uesaka2016,Uesaka2018}:
\begin{align}
\frac{d^2\Gamma_{unpol.}}{d\epsilon_1dc_{12}}=& \sum_{l\ge 0}c_l\left(\epsilon_1\right)P_l(c_{12}),
\label{eq:dif_rate}
\end{align}
where $P_l(x)$ is the Legendre polynomial and the coefficient $c_l$ is
\begin{align}
c_l\left(\epsilon_1\right)=&\left(E_{tot}-2m_e\right)\frac{G_F^2}{2\pi^3}|\bm{p}_1||\bm{p}_2|\left(2j_{\kappa_e}+1\right) \nonumber\\
&\times\sum_{\kappa_1,\kappa_2}\sum_{\kappa'_1,\kappa'_2}\sum_{J}(-1)^{J-j_{\kappa_2}-j_{\kappa'_2}}\frac{1+(-1)^{l_{\kappa_1}+l_{\kappa'_1}+l}}{2}\frac{1+(-1)^{l_{\kappa_2}+l_{\kappa'_2}+l}}{2} \nonumber\\
&\times i^{-l_{\kappa_1}-l_{\kappa_2}+l_{\kappa_1'}+l_{\kappa_2'}}e^{i\left(\delta_{\kappa_1}+\delta_{\kappa_2}-\delta_{\kappa_1'}-\delta_{\kappa_2'}\right)}N(J,\kappa_1,\kappa_2,E_1,\alpha_e)N^*(J,\kappa'_1,\kappa'_2,E_1,\alpha_e) \nonumber\\
&\times(2J+1)\left(2j_{\kappa_1}+1\right)\left(2j_{\kappa_2}+1\right)\left(2j_{\kappa_1'}+1\right)\left(2j_{\kappa_2'}+1\right) \nonumber\\
&\times\left(2l+1\right)
\begin{pmatrix}
j_{\kappa_1} & j_{\kappa'_1} & l \\
1/2 & -1/2 & 0
\end{pmatrix}
\begin{pmatrix}
j_{\kappa_2} & j_{\kappa'_2} & l \\
1/2 & -1/2 & 0
\end{pmatrix}
\begin{Bmatrix}
j_{\kappa_1} & j_{\kappa_2} & J \\
j_{\kappa'_2} & j_{\kappa'_1} & l
\end{Bmatrix}.
\end{align}
The $N$ is represented by
\begin{align}
N\left(J,\kappa_1,\kappa_2,E_1,\alpha_e\right)=N_\mathrm{photo}+N_\mathrm{contact},
\end{align}
with
\begin{align}
N_\mathrm{photo}=&\sum_{j=L/R}A_jW_j(J,\kappa_1,\kappa_2,E_1,\alpha_e) \\
N_\mathrm{contact}=&\sum_{j=1}^{6}g_jW_j(J,\kappa_1,\kappa_2,E_1,\alpha_e).
\end{align}

Here the $W_j$'s ($j=1,2,\cdots,6$) for the contact interaction are given by
\begin{align}
W_1(J)=&\frac{1}{2}\left\{X^-_\alpha(J,0,J)-X^+_\beta(J,0,J)+i\left[Y^+_\alpha(J,0,J)+Y^+_\beta(J,0,J)\right]\right\}, \\
W_2(J)=&\frac{1}{2}\left\{X^-_\alpha(J,0,J)-X^+_\beta(J,0,J)-i\left[Y^+_\alpha(J,0,J)+Y^+_\beta(J,0,J)\right]\right\}, \\
W_3(J)=& 2\left\{X^-_\alpha(J,0,J)+X^+_\beta(J,0,J)-i\left[Y^+_\alpha(J,0,J)-Y^+_\beta(J,0,J)\right]\right\}, \\
W_4(J)=& 2\left\{X^-_\alpha(J,0,J)+X^+_\beta(J,0,J)+i\left[Y^+_\alpha(J,0,J)-Y^+_\beta(J,0,J)\right]\right\}, \\
W_5(J)=& 3\sum_{L=|J-1|}^{J+1}X^-_\beta(L,1,J)-X^+_\alpha(J,0,J)+i\left[3\sum_{L=|J-1|}^{J+1}Y^-_\alpha(L,1,J)+Y^-_\beta(J,0,J)\right], \\
W_6(J)=& 3\sum_{L=|J-1|}^{J+1}X^-_\beta(L,1,J)-X^+_\alpha(J,0,J)-i\left[3\sum_{L=|J-1|}^{J+1}Y^-_\alpha(L,1,J)+Y^-_\beta(J,0,J)\right],
\end{align}
with
\begin{align}
X^\pm_\alpha(L,S,J)=& Z_{gggg}(L,S,J)+Z_{ffff}(L,S,J)\pm\left[Z_{gfgf}(L,S,J)+Z_{fgfg}(L,S,J)\right], \\
X^\pm_\beta(L,S,J)=& Z_{ggff}(L,S,J)+Z_{ffgg}(L,S,J)\pm\left[Z_{gffg}(L,S,J)+Z_{fggf}(L,S,J)\right], \\
Y^\pm_\alpha(L,S,J)=& Z_{ggfg}(L,S,J)-Z_{ffgf}(L,S,J)\pm\left[Z_{fggg}(L,S,J)-Z_{gfff}(L,S,J)\right], \\
Y^\pm_\beta(L,S,J)=& Z_{gggf}(L,S,J)-Z_{fffg}(L,S,J)\pm\left[Z_{gfgg}(L,S,J)-Z_{fgff}(L,S,J)\right],
\end{align}
and
\begin{align}
Z_{ABCD}(L,S,J)=& \int_{0}^{\infty}drr^2A_{p_1}^{\kappa_1}(r)B_{1,\mu}^{\kappa_\mu}(r)C_{p_2}^{\kappa_2}(r)D_{n,e}^{\kappa_e}(r) \nonumber\\
&\times\sqrt{\left(2l^A_{\kappa_1}+1\right)\left(2l^B_{\kappa_\mu}+1\right)\left(2l^C_{\kappa_2}+1\right)\left(2l^D_{\kappa_e}+1\right)} \nonumber\\
&\times\left(l^A_{\kappa_1},0,l^C_{\kappa_2},0|L,0\right)\left(l^B_{\kappa_\mu},0,l^D_{\kappa_e},0|L,0\right) \nonumber\\
&\times
\begin{Bmatrix}
l^A_{\kappa_1} & 1/2 & j_{\kappa_1} \\
l^C_{\kappa_2} & 1/2 & j_{\kappa_2} \\
L & S & J
\end{Bmatrix}
\begin{Bmatrix}
l^B_{\kappa_\mu} & 1/2 & j_{\kappa_\mu} \\
l^D_{\kappa_e} & 1/2 & j_{\kappa_e} \\
L & S & J
\end{Bmatrix},
\end{align}
where $\kappa_\mu=-1$.
Here $A$ and $C$ represent the electron scattering states with momenta $p_1$ and $p_2$, and $B$ and $D$ represent the bound states of the muon and electron.
The radial wave functions $A(r)$, $B(r)$, $C(r)$, and $D(r)$ are either $g(r)$ or $f(r)$.
The angular momentum $l^h_\kappa$ is defined as
\begin{align}
l^h_\kappa=
\begin{cases}
l_{+\kappa} & \ \mbox{for} \ \ h=g, \\
l_{-\kappa} & \ \mbox{for} \ \ h=f.
\end{cases}
\end{align}

The amplitudes of the photonic interaction $W_{L/R}$ are given as
\begin{align}
W_{L/R}=&\frac{2m_\mu}{i}\sqrt{\pi\alpha}\sum_{l=0}^{\infty}\sum_{j=|l-1|}^{l+1}\sum_{\lambda=1}^{3}\left[X_\lambda\left(l,j,\kappa_1,\kappa_2,J\right)\pm iY_\lambda\left(l,j,\kappa_1,\kappa_2,J\right)\right],
\label{eq:W_LR}
\end{align}
where $\pm$ corresponds to $L$ and $R$, respectively.
$X_\lambda$ and $Y_\lambda$ are expressed in terms of $Z$ as
\begin{align}
X_1\left(l,j,\kappa_1,\kappa_2,J\right)=&(-1)^{l+j}\left\{Z_{gfgf}^{l,l,1,j}(J)+Z_{fggf}^{l,l,1,j}(J)-Z_{gffg}^{l,l,1,j}(J)-Z_{fgfg}^{l,l,1,j}(J)\right\}, \\
X_2\left(l,j,\kappa_1,\kappa_2,J\right)=& f_{l-j}^{(2)}(j)\left\{Z_{gfgg}^{l,j,0,j}(J)+Z_{fggg}^{l,j,0,j}(J)+Z_{gfff}^{l,j,0,j}(J)+Z_{fgff}^{l,j,0,j}(J)\right\}, \\
X_3\left(l,j,\kappa_1,\kappa_2,J\right)=& f_{l-j}^{(3)}(j)\sum_{\{l_a,l_b\}=\{l,j\},\{j,l\}}\left\{Z_{gggf}^{l_a,l_b,1,j}(J)-Z_{ffgf}^{l_a,l_b,1,j}(J)-Z_{ggfg}^{l_a,l_b,1,j}(J)+Z_{fffg}^{l_a,l_b,1,j}(J)\right\},
\end{align}
\begin{align}
Y_1\left(l,j,\kappa_1,\kappa_2,J\right)=&(-1)^{l+j}\left\{Z_{gggf}^{l,l,1,j}(J)-Z_{ffgf}^{l,l,1,j}(J)-Z_{ggfg}^{l,l,1,j}(J)+Z_{fffg}^{l,l,1,j}(J)\right\}, \\
Y_2\left(l,j,\kappa_1,\kappa_2,J\right)=& f_{l-j}^{(2)}(j)\left\{Z_{gggg}^{l,j,0,j}(J)-Z_{ffgg}^{l,j,0,j}(J)+Z_{ggff}^{l,j,0,j}(J)-Z_{ffff}^{l,j,0,j}(J)\right\}, \\
Y_3\left(l,j,\kappa_1,\kappa_2,J\right)=& f_{l-j}^{(3)}(j)\sum_{\{l_a,l_b\}=\{l,j\},\{j,l\}}\left\{Z_{gffg}^{l_a,l_b,1,j}(J)+Z_{fgfg}^{l_a,l_b,1,j}(J)-Z_{gfgf}^{l_a,l_b,1,j}(J)-Z_{fggf}^{l_a,l_b,1,j}(J)\right\},
\end{align}
where
\begin{align}
f_{h}^{(2)}(j)=
\begin{cases}
\sqrt{\dfrac{j+1}{2j+1}} & (h=+1) \\
0 & (h=0) \\
\sqrt{\dfrac{j}{2j+1}} & (h=-1)
\end{cases}, \hspace{10mm}
f_{h}^{(3)}(j)=
\begin{cases}
\sqrt{\dfrac{j}{2j+1}} & (h=+1) \\
0 & (h=0) \\
-\sqrt{\dfrac{j+1}{2j+1}} & (h=-1)
\end{cases}.
\end{align}
The matrix element $Z$, which consists of the CLFV and the electromagnetic vertex and
the photon propagator, is given by
\begin{align}
Z_{ABCD}^{l_a,l_b,s,j}(J)=&\left[q_0^2\int_{0}^{\infty}dr_1r_1^2A_{p_1}^{\kappa_1}(r_1)B_{1,\mu}^{\kappa_\mu}(r_1)\int_{0}^{\infty}dr_2r_2^2F_{l_a,l_b}^{q_0}(r_1,r_2)C_{p_2}^{\kappa_2}(r_2)D_{n,e}^{\kappa_e}(r_2)\right. \nonumber\\
&\left.\times(-1)^{J+\kappa_1+\kappa_\mu}V_{l_a,1,j}^{s_A\kappa_1,s_B\kappa_\mu}V_{l_b,s,j}^{s_C\kappa_2,s_D\kappa_e}
\begin{Bmatrix}
j_{\kappa_1} & j_{\kappa_2} & J \\
j_{\kappa_e} & j_{\kappa_\mu} & j
\end{Bmatrix}\right] \nonumber\\
&-(-1)^{j_{\kappa_1}+j_{\kappa_2}-J}\left[\left\{p_1,\kappa_1\right\}\leftrightarrow\left\{p_2,\kappa_2\right\}\right],
\label{eq:Z_ABCD}
\end{align}
where $\kappa_\mu=-1$ again.
$q_0$ is $q_0=m_\mu-B_\mu^{1s}-E_1$ for the direct term,
and $q_0=m_\mu-B_\mu^{1s}-E_2$ for the exchange term.
The coefficients $V$ are given by
\begin{align}
V_{l,s,j}^{\kappa_b,\kappa_a}=&(-1)^{l}\frac{1+(-1)^{l_{\kappa_b}+l_{\kappa_a}+l}}{2}\left(j_{\kappa_b},1/2,j_{\kappa_a},-1/2|j,0\right) \nonumber\\
&\times
\begin{cases}
\delta_{l,j} & (s=0, j=l) \\
(j-\kappa_a-\kappa_b)/\sqrt{j(2j+1)} & (s=1, j=l+1) \\
(\kappa_a-\kappa_b)/\sqrt{j(j+1)} & (s=1, j=l) \\
-(j+1+\kappa_a+\kappa_b)/\sqrt{(j+1)(2j+1)} & (s=1, j=l-1)
\end{cases}.
\end{align}
We use $s_A = \pm 1$ for $A=g$ and $A=f$, respectively.
$F_{l_a,l_b}^{q_0}(r_1,r_2)$ is the coefficient in the partial wave expansion of the photon propagator, which is given as
\begin{align}
\int\frac{d^3q}{(2\pi)^3}\frac{q_\nu e^{-i\bm{q}\cdot\left(\bm{r}_1-\bm{r}_2\right)}}{|\bm{q}|^2-q_0^2-i\epsilon}=q_0\partial_\nu\sum_{l,m}Y_{l}^{m*}\left(\hat{r}_1\right)Y_{l}^{m}\left(\hat{r}_2\right)F_{l,l}^{q_0}\left(r_1,r_2\right),
\end{align}
where we have defined $\partial_{\nu}=(iq_0,\bm{\nabla}_1)$ and
\begin{align}
F_{l_1,l_2}^{q_0}\left(r_1,r_2\right)=h_{l_1}^{(1)}\left(q_0r_1\right)j_{l_2}\left(q_0r_2\right)\theta(r_1-r_2)+h_{l_2}^{(1)}\left(q_0r_2\right)j_{l_1}\left(q_0r_1\right)\theta(r_2-r_1).
\end{align}
Here $j_l$ and $h_l^{(1)}$ are the spherical Bessel function and the first kind
spherical Hankel function, respectively.

\end{document}